\documentclass[12pt]{article}

\def\ApJ{\sl Astrophys. J.}
\def\MNRAS{\sl Mon. Not. R. Astron. Soc.}
\def\AA{\sl Astron. Astrophys}

\title{\bf {Gravitational Lensing Bound On The Average Redshift Of Gamma Ray Bursts In Models With Evolving Lenses }}
\author{Deepak Jain\thanks{E--mail : deepak@ducos.ernet.in},
N.Panchapakesan\thanks{E--mail : panchu@ducos.ernet.in},
S.Mahajan\thanks{E--mail : sm@ducos.ernet.in} and 
V.B.Bhatia\thanks{E--mail : vbb@ducos.ernet.in} \\
	{\em Department of Physics and Astrophysics} \\
        {\em University of Delhi, Delhi-110 007, India} 
	}

\usepackage{overcite}

\begin {document}
\maketitle

\begin{center}
\Large{\bf Abstract}
\end{center}
\large
\baselineskip=20pt

Identification  of gravitationally lensed Gamma Ray Bursts
(GRBs) in the BATSE 4B catalog can be used to constrain the average
redshift $ < z > $ of the GRBs. In this paper we investigate the
effect of evolving lenses on the $< z >$ of GRBs in
different cosmological models of universe. The cosmological parameters
$\Omega$ and $\Lambda$ have an effect on the $ < z >$ of GRBs. The
other factor which can change the $ < z > $ is the evolution of
galaxies. We consider three evolutionary model of galaxies. In
particular, we find that
the upper limit on $ < z >$ of GRBs is higher in evolving model of
galaxies as compared to non-evolving models of galaxies.

      \vfill
      \eject

\large

\baselineskip = 20pt
\begin {section} {Introduction}
The use of gravitational lensing as a tool for the determination of
cosmological parameters (e.g. $H_{0}, \Omega_{0},
\Lambda_{0}$)  has frequently been discussed \cite{ref, press, f1, t}.
To constrain these
parameters either QSOs or galaxies have been used as sources. To use
Gamma Ray Bursts (GRBs) as a source for gravitational lensing is not a
new idea. As first discussed by Paczynski \cite{pac1, pac2}, if GRBs are
cosmological then they should be gravitationally lensed just as quasars.

There is now overwhelming evidence that the majority of GRB sources
lie at cosmological distances. The detection of GRB990123 which is
believed to lie between redshift\cite{jens}  $1.6 \leq z < 2.14$, the
identification of the host galaxy\cite{kul} for  GRB981214 at $ z = 3.42$ 
and the detection of afterglows of many others GRBs\cite {costa, pir}
in X-rays, optical and  radio domain by
BeppoSAX satellite are all suggestive of GRBs being at cosmological
distances. Therefore it is expected that
the GRBs  must be
gravitationally lensed as the probability of lensing increases with
the source redshift.
The primary effect of a gravitational lens on GRB would be to create
more than one image of the burst. These images could not be angularly
resolved with the present technology since BATSE's spatial resolution is
insufficient to resolve the images angularly but they could be
temporally resolved. Several authors \cite{bla,go,nemi1,mar} have
presented detailed
calculations of GRB lensing which can be used as a probe in the search
for dark matter in the form of compact objects.

So far no lensed GRBs have been detected with BATSE which clearly 
implies an
upper limit to the average redshift $ < z >$ of GRBs as first
calculated by Holz, Miller \& Quashnock \cite{holz} [HMQ]. HMQ
calculated an upper
limit to $ < z >$  which is independent of the physical model of GRBs. 
They have calculated $ < z >$ for GRBs in different cosmological models.

Till date a substantial fraction of all GRBs observed occur at
significant cosmological distances ( $ z \sim 0.8 - 3.4$). Further the
`` no host galaxy problem'' pushes the GRBs to be, either at very high
redshift ( $z > 6$) or not to be in normal host galaxies \cite{sch}. Several authors\cite{wij,tot} propose that the GRBs rate should trace the star
formation rate in
the universe and consequently it places the very dim burst\cite{wij}
at $ z \geq6$. HMQ calculated the $ < z >$ of GRBs in a non-evolving model of
galaxies (lenses) and the $ < z > $ at 95 $\%$ confidence level (CL)
comes out to be $ <
2.2, 2.8, 4.3 $ for different combinations of ($\Omega$, $\Lambda$)
namely (0.3, 0.7), (0.5, 0.5) and (0.5, 0.0) respectively.
These results get considerably 
modified when {\it evolution in the properties of lensing galaxies
(lenses)} is considered. The modified results permit a larger value of
$ < z >$ and are thus consistent to some extent with the high - z GRB
scenarios.
The purpose of this paper is to put the 
upper limit on $< z >$ of GRBs in different models of evolving 
lensing galaxies.

We refine the analysis of HMQ by considering the evolution of galaxies.
Normally the comoving number density of galaxies is assumed to be
constant while calculating lensing probability. But it is an
oversimplification to assume that galaxies are formed at a single
epoch. Evolution tells 
us how the mass  or number density of the lens varies with
cosmic time scales. Merging between galaxies and the infall of
surrounding  mass into galaxies are two possible processes that can
change the comoving density of galaxies and/or their mass. The effect
of galaxy merging or evolution has been studied by many authors
\cite{m,m1,rix,j1,j2}.  Most of them have  focused on
the statistical properties of gravitational lenses  and the limits on
the  cosmological constant.
This work is an attempt to check  how galaxy evolution 
changes the upper limit on the average redshift of GRBs.

The structure of the paper is as follows. In section 2 we describe 
the evolutionary model of galaxies that we use. The lensing probability with
evolving lenses is given in section 3. In section 4 we put  the
upper limits on the average redshift of GRBs, assuming that no
lensing events are present in the BATSE 4B catalog.
Results on $< z >$ of GRBs with evolving lensing galaxies in different
cosmological models are described in section 5.
Discussion is given in section 6.

\end {section}

\begin {section} {Evolution Of Galaxies}

\noindent Galaxy mergers and gravitational lensing are interlinked with
each other as the merging of small galaxies gives rise to elliptical
galaxies which further act as gravitational lenses. We consider 
three evolutionary models of galaxies 
\vskip 0.3cm
\noindent {\bf{\underline{Fast Merging}}}
\vskip 0.3cm
This  evolutionary model was proposed by
Broadhurst, Ellis \& Glazebrook (1992) [BEG] \cite{bro} in order  
to resolve
the faint galaxy population counts. The basic point of this model is
to introduce strong number evolution  or to assume that galaxy numbers are
not conserved with cosmic time but
the total comoving mass remains conserved. This model assumes the
comoving number
density of the lenses to vary with cosmic time as : 

\begin{equation}
n(\delta t) = f(\delta t) n_{0}
\end{equation}

\noindent where $\delta t $ is the look-back time and the subscript
$'0'$ indicates the present day values. The characteristic mass ( or
dispersion velocity ) for
self similar galaxy mass function  of
Singular Isothermal Sphere (SIS) lenses at  $\delta $t is

\begin{equation}
v(\delta t) = [f(\delta t)]^{-1} v_{0}
\end{equation}
This model is better than the model proposed by Volmerange and
Guiderdoni \cite{vol} where the merging goes exponentially with $( 1 + z)$,
because the merging rate doesn't become high at early time as
explained by BEG.
According to the BEG model  if we had $n$ galaxies at look-back time
$\delta t$ each with velocity dispersion $v$, they would by today have
merged into one galaxy with a velocity dispersion $[f(\delta t)]v
$. The function $f(\delta t)$ describes the time dependence and the  
strength of merging :

\begin{equation}
f(\delta t) = exp(Q H_{0}\delta t)
\end{equation}

\noindent where $H_{0}$ is the Hubble constant at the present epoch
and Q represents 
the merging 
rate. We take Q = 4 as suggested by BEG \cite{bro}. The 
look back time $\delta t$ is related to $z_{1}$ through

\begin{equation}
H_{0}\delta t = \int^{z_{1}}_{0}{(1+y)^{-1} dy \over{\sqrt{F(y)}}}
\end{equation}

\noindent where  $F(y)=\Omega_{0}(1 + y)^{3}+(1-\Omega_{0}-\Lambda_{0})(1 +
y)^{2}+\Lambda_{0}$ 

\noindent where $\Omega_{0} = 8 \pi G \rho_{0} /3 H_{0}^2$ ,
$\rho_{0}$ is density
of matter, $\Lambda_{0} = 8 \pi G \rho_{v0} /3 H_{0}^2$ and 
$\rho_{v0}$  is density
of vacuum 

\vskip 1cm

\noindent{\bf{\underline {Slow Merging}}}
\vskip 0.5cm
\noindent In this slow  merging  model\cite{gun} the mass of an
individual galaxy increases with cosmic time as $t^{2/3}$ while
the comoving
number density varies with cosmic time as $t^{-2/3}$, 
so the total mass of the
galaxies within a given comoving volume is conserved 
The cosmic time  $t$ starts from the big bang. We also assume that the
power law relation between the mass and velocity
is  $ M \propto v^{3.3} $ for elliptical galaxies
 \cite{naka}. Then

\begin{equation}
n(\delta t) =  n_{0}\left[1 - {\delta t \over t_{0}}\right]^{-2/3}
\end{equation}

\begin{equation}
v(\delta t) = v_{0}\left[1 - {\delta t \over t_{0}}\right]^{1/5}
\end{equation}

\noindent where $t_{0}$ is present age of the universe.

\vskip 1cm

\noindent{\bf{\underline{Mass Accretion}}}

\vskip 0.5cm

\noindent Mass accretion is the key factor for evolution of galaxies. 
In this model\cite{park} the comoving
number density of  the galaxies remains
constant but the mass of galaxy increases with cosmic time as $t^{2/3}$
in the same way as in the slow merging model.

\begin{equation}
n(\delta t) =  n_{0}(constant)
\end{equation}

\begin{equation}
v(\delta t) = v_{0}\left[1 - {\delta t \over t_{0}}\right]^{1/5}
\end{equation}

\end{section}

\begin{section} {Basic Equations For Gravitational Lensing Statistics}
\noindent
The differential probability $d\tau$ of a beam encountering a lens 
in traversing the path of $dz_{L}$ is given by\cite{TOG, f}

\begin{equation} d\tau = n_{L}(z)\sigma{cdt\over dz_{L}}
dz_{L},\end{equation}.

\noindent
 where $n_{L}(z)$ is 
the comoving number density.

The Singular Isothermal Sphere (SIS) provides us with a reasonable
approximation to account for the lensing properties of a real 
galaxy. The lens model
is characterized
by the one dimensional velocity dispersion $v$. The deflection
angle for all impact parameters is given by $\tilde{\alpha} =
 4\pi v^{2}/c^{2}$. This lens produces two images if the angular
position of the source is less than the critical angle $\beta_{cr}$,
which is the deflection of a beam passing  at any radius through a SIS:

\begin{equation} \beta_{cr} =\tilde{\alpha} D_{LS}/D_{OS} ,\end{equation}

\noindent Here we use the notation $D_{OL}=d(0,z_{L}), D_{LS} =d(z_{L},z_{S}), D_{OS} =d(0,z_{S})$, where $d(z_{1}, z_{2})$
is the angular diameter distance between the redshift $z_{1}$ and
$z_{2}$  \cite{f}.
\noindent
Then the critical impact parameter is defined by $a_{cr} = 
D_{OL}\beta_{cr}$ and the cross- section is given by 

\begin{equation}
\sigma = \pi a_{cr}^{2} = 16{\pi}^{3}\left({v \over c}\right)^{4}
\left({D_{OL}D_{LS}\over D_{OS}}\right)^{2} ,
\end{equation}

\noindent
\vskip 0.3in

\noindent{\bf {\underline{ The Evolutionary Model}}}

\noindent The differential probability $d\tau$ of a lensing event in an
evolutionary model 
can be written as:
\begin{eqnarray}
{d\tau}&=& {16\pi^{3}\over{c
H_{0}^{3}}}\,\phi_\ast\, v_\ast^{4}\Gamma\left(\alpha +
{4\over\gamma}+1 \right) f(\delta t)^{( 1 - {4\over\gamma})} \nonumber\\
& & \nonumber\\
& &\times\,(1 + z_{L})^{3}\left({D_{OL}D_{LS}\over R_\circ
D_{OS}}\right)^{2}{1\over R_\circ}{cdt  \over dz_{L}} dz_{L}
\end{eqnarray} 

$${d\tau} = F(1 + z_{L})^{3}\left({D_{OL}D_{LS}\over R_\circ
D_{OS}}\right)^{2}f(\delta t)^{( 1 - {4\over\gamma})}
{1\over R_\circ}{cdt \over dz_{L}} dz_{L}$$.

\noindent where  $ F ={16\pi^{3}\over{c
H_{0}^{3}}}\,\phi_\ast\, v_\ast^{4}\Gamma \left(\alpha + {4\over\gamma}
+1 \right)$.
By substituting the values of $\phi_\ast, \alpha$ as given by Corray
et al. \cite{cor} and
$v_\ast, \gamma$ as given by Nakamura \& Suto \cite{naka} we get $ F =
0.035$. 
\noindent
where $ f(\delta t) = exp(Q H_{0} \delta t)$ for fast merging
and $ f(\delta t) = \left(1 - {\delta t \over t_{0}}\right)^{-2/3}$
for slow 
merging. In case of mass accretion $ f(\delta t) = \left(1 - {\delta t
\over t _{0}}\right)^{-2/3}$ but the  exponent of $ f(\delta t)$ for
mass accretion model in
eq. (12)  becomes
 $(-1 -{ 4\over \gamma})$ as the total mass in galaxies increases with
time.

\noindent

\vskip .3in
\noindent {\bf {\underline{ The Non Evolutionary Model}}}

\noindent
In the non merging model the optical depth is given by \cite{f}

\begin{eqnarray}
{d\tau} &=& {16\pi^{3}\over{c
H_{0}^{3}}}\,\phi_\ast\, v_\ast^{4}\Gamma\left(\alpha + {4\over\gamma}
+1\right)(1 + z_{L})^{3} \nonumber\\
& & \nonumber\\
&&\times\,
\left({D_{OL}D_{LS}\over R_\circ D_{OS}}\right)^{2}{1\over
R_\circ}{cdt \over  dz_{L}} dz_{L}
\end{eqnarray} 

\end{section}

\begin{section}{Bound On Average Redshift Of GRBs }

\vskip 0.2cm

\noindent Adopting the simplest matter distribution for the lensing
galaxy as a Singular Isothermal Sphere (SIS), we calculate the
probability $\tau(z)$ of a beam from a source at redshift z is imaged
by a lens in the filled beam approximation.
Apart from the cosmological parameters $\Omega$ and $\Lambda$, the lensing
probability depends not only on the model used for describing the
evolution of galaxies but most importantly on the parameter
$F$\cite{TOG,ft,f,koch}. $F$ parameterizes the distribution of galaxies
as well as their effectiveness in the lensing process (defined in
section 3). Thus the elliptical
galaxies contribute strongly while spirals have a negligible effect. The
Schechter distribution function often forms the basis of these
estimates. However it is being increasingly realised that
sub-distribution of specific types of galaxies have to be further taken
into account. We take F = 0.035
which lies in the current estimate range\cite{j1,chi,cor,cor1}  
$0.02 < F < 0.05$.
In the SIS
model, the lensing event always consists of two images and the third
central image is too faint to be observed. We take a constant BATSE
efficiency \cite{hak}  $\epsilon = 0.48$  to see either
image and ${\epsilon}^{2} = 0.23$ to see both images. So far no
multiple images has been detected, which may due to the low
instrumental  efficiency.

In order to put the bound on the $< z > $ of GRBs, we follow the
methodology of HMQ. As the distribution of GRBs in redshift is
unknown, we use the simplifying assumption that all the sources are
at the same redshift. We estimate the number of image pairs as  

\begin{equation}
N_{< z >} = (N_{tot}/\epsilon).{\epsilon}^{2}.\tau( < z >) =
N_{tot}\epsilon\tau(< z >)
\end{equation}

where $N_{tot}$ be the total number of observed
bursts in  the BATSE 4B catalog (which are 1802)
then $N_{tot}/\epsilon$ are actual burst sources above the
BATSE threshold. The $ < z >$ in the above equation tells the average
redshift of GRBs, assuming that all the sources are at this
redshift. 
If we define $\phi_{min}$ as the brightness
threshold  below which identification of lensed images from light-curve
comparison is impossible, then the expected number gets modified to
\cite{holz,gross}

\begin{equation}
N_{< z >} =
N_{tot} \epsilon\tau(< z >)\int_{0}^{\infty}dB(\phi)
{1 \over {[1 + \phi_{min}/\phi(z)]^{2}}}
\end{equation}

We plot this $N_{<z>}$ with $ < z > $ in different cosmological models
as shown in Fig. 1, Fig.2 and Fig.3 respectively.
Here $ B(\phi)$ is the  the observed  BATSE
brightness distribution and the integrand is the conditional
probability that the both images are above the brightness threshold.
The value of this integral is equal to 0.57 \cite{holz}. 

   \end {section}

\begin{section}{Results}
We have taken three representative values of $(\Omega,\Lambda)$ as
(0.3, 0.7), (0.5, 0.5) and (0.2, 0). With these values we calculate the
expected number of observable image pairs in the BATSE 4B as a
function of average redshift for the mass accretion, fast merging,
slow merging and no evolutionary models.  We find that
the best limit arises with a large cosmological constant where the
lensing rate is quite high. At the 95 $\%$ confidence level, we find
 an upper limit on $\left < z \right > < 4.3, \ 7.8,\ 9.8 $ 
in the non-evolutionary 
model for $(\Omega,\Lambda)$ values of
(0.3, 0.7), (0.5, 0.5) or (0.2, 0) respectively. The upper limit on
$\left < z \right > $ at 95$\%$ CL in other evolutionary models are 
described in table 1.
At 68 $\%$ confidence level, the slow merging model gives  
upper limit on $\left< z\right >  < 2.3,\ 3.3 \ or \ 3.9 $ for 
$(\Omega,\Lambda)$ values of
(0.3, 0.7), (0.5, 0.5) or (0.2, 0) respectively. Similarly fast merging
model gives $\left < z \right> < 2.9, \ 4.3 \ or \ 5.2 $ for 
$(\Omega,\Lambda)$ values of
(0.3, 0.7), (0.5, 0.5) or (0.2, 0 ) respectively as shown in fig.1, 
fig.2, and fig.3 respectively. The mass accretion model at 68$\%$ CL 
gives an upper limit on $\left< z\right > < 6.1$ only for
 $(\Omega = 0.3, \Lambda = 0.7)$.
The result is described in Table 1 and Table 2.

\end{section}

\begin{section} {Discussion}

\noindent We notice that the evolution of lensing galaxies serves to
increase the $ < z > $ of GRBs as compare to non- evolutionary model
of galaxies.

\noindent Throughout the paper, we have used
the filled beam approximation (standard distance) \cite{TOG,f} in
which ``smoothness parameter'' $\alpha$ ( which measures the degree of
inhomogeneity of the universe) is equal to $1$. On the other hand
with $\alpha = 0$ one gets empty beam distance in which there is negligible
intergalactic matter and the line of sight to a distant object doesn't
pass close to intervening galaxies\cite{dr1,dr2}. At present we don't
know what value of smoothness parameter in the distance formula
describes our universe best. Ehlers and Schneider \cite{es}(1986)
argued that the direction to the source cannot be a random variable in the
clumpy universe. 
At lower redshift the probability of lensing in different distance
formulations remains approximately the same while at higher redshift
the lensing rate become different in different distance approximations.
So the question arises, which is the best distance approximation?
But recently in an
elegant paper, Holz \& Wald (1998) \cite{holz1} have exhibited a new
method  for
calculating  gravitational lensing rates. This method is free from the
ambiguities present in the distance formula.
In this formulation, the optical depth at any redshift
is greater than the optical depth calculated using the angular diameter
distance in the filled beam approximation.
Thus we expect that this formulation will increase the expected
lensing rate, and hence decrease the upper limit on $<z>$ of GRBs. 

Another reason why our lensing rate could be an underestimate because
of the assumption that lensing is only due to the SIS galaxies. But
there are other 
lens models also like: point like mass as deflector,
isothermal sphere with a
softened core, asymmetric lens , spherical  model (e.g. King model),
cosmic strings and constant density sheet which in principle could also
contribute. Many authors \cite{ mao,nemi1,nemi2,nemi3,ma2}
have explored the possibility of GRBs lensing by point masses. Grossman
and Nowak \cite{gross} also studied the lensing of GRBs by asymmetric
and non singular isothermal spheres lenses. These
additional lensing objects may change $< z >$.

\vskip 0.5cm
\noindent 
\vskip 0.5cm
\noindent The overall probability of strong lensing depends directly 
on the parameter
$F$, which further depends upon four parameters $\alpha,
\gamma,\\ 
v_{\ast}, \phi_{\ast}$.
There are many uncertainties
associated with
these parameters as mentioned in sec. 4 and discussed by several
authors \cite{k96,chi,chen,cor}. These uncertainties
can change the 
value of $F$ by upto $30\%$  and hence  affect our results.
HMQ used the value of $F = 0.1$ which is nearly three times higher
than the value of F used in this paper. Therefore the upper limit on
the $< z >$ of GRBs in HMQ paper is less than our value of $< z >$.

\vskip 1cm

The ability of BeppoSax satellite to detect GRBs opens up a new
era in the studies of GRBs. Many counterparts of GRBs have been detected in
X-ray, optical and radio domains from which important insight in
this field has been gained.  
Nevertheless, the nature of
the  central engine  that 
accelerates the relativistic flow is still not very clear. In order to
calculate the amount of total energy in the burst, the precise
measurement  of redshift is required. For example, spectroscopic
observations show that the host galaxy for GRB971214 is at a redshift
 z = 3.418. Given this high redshift and the known fluence of this GRB
the $\gamma$ - ray energy release
of this burst is unexpectedly large, about $3 \times 10^{53} erg$,
assuming isotropic emission  \cite{kul}.
Energy released in other forms of
radiation is not included in this energy calculation. The currently
favored model for GRBs is coalescence of neutron stars
\cite{nar}. The coalescence model is expected to released about
$10^{51} erg$ in the form of
electromagnetic energy. So there is big gap between the model based
calculation and the parameters based calculation.
Similarly,  with the combination of a redshift $\geq 1.6$ and fluence
for GRB990123 \cite{kul1} imply that $\gamma$ ray energy release in
this  burst is
$ 3.4 \times 10^{54} erg$, assuming the emission is isotropic. This would
again put strain on any GRB model based on merging of neutrons star or black
holes. At higher redshift this problem becomes more severe. We may be
forced to consider even more energetic possibilities\cite{pac0} or to
find ways of
extracting more electromagnetic energy in coalescence models or we
have to search for ways to reduce the estimated energy\cite{kul1,mich}
release from GRBs to resolve this problem.

\end{section}

\begin{section}*{Acknowledgements}

We thank Myeong - Gu Park, D. Holz, M. Coleman Miller, Lawrence
M. Krauss, M. Chiba, T. T. Nakamura and Yu - Chung N. Cheng for useful
discussions. 

\end {section}

\vfill
\eject

\begin{figure}[ht]
\vskip 15 truecm
\includegraphics{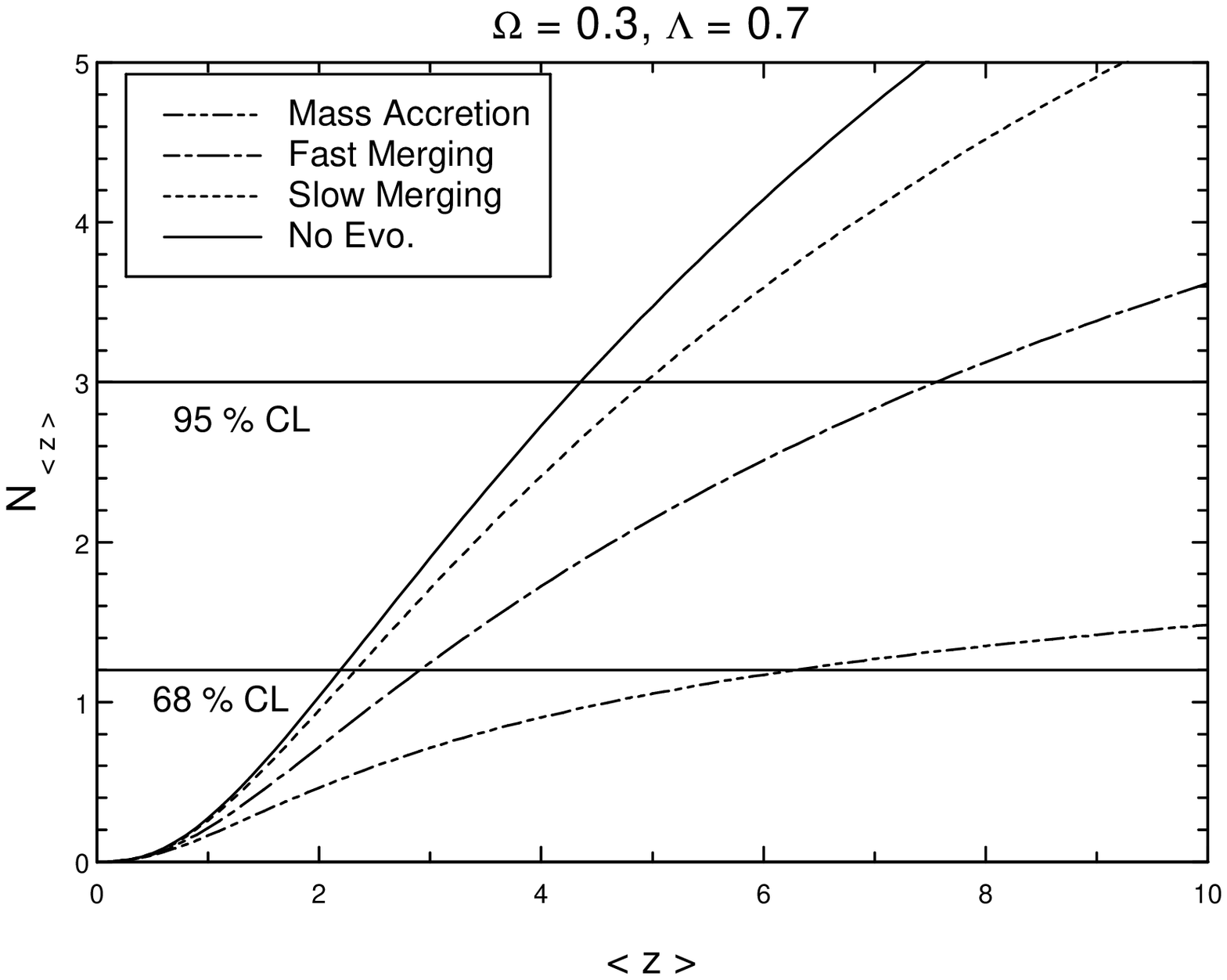}
\caption{}

\end{figure}

\begin{figure}[ht]

\vskip 15 truecm
\includegraphics{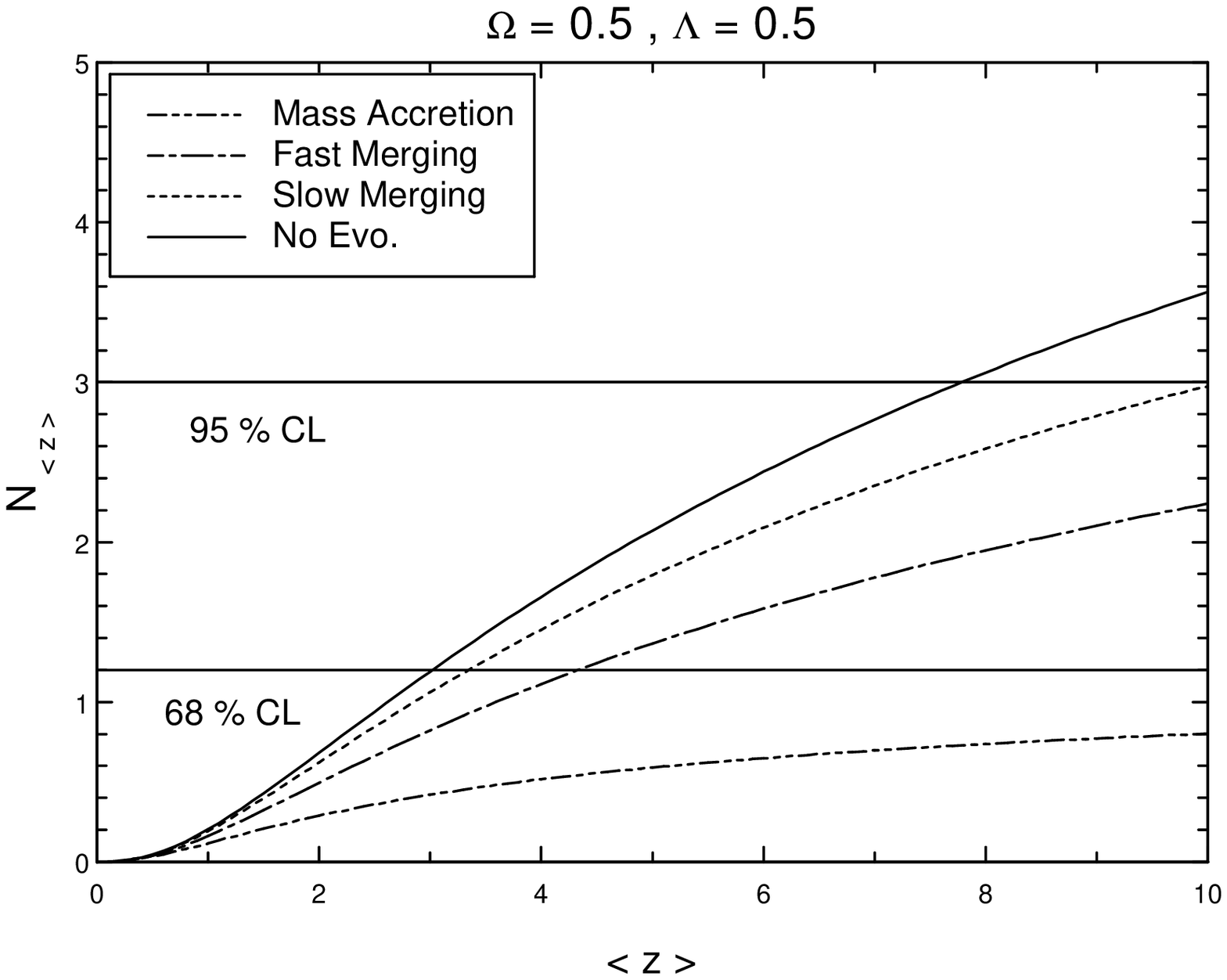}
\caption{}

\end{figure}

\begin{figure}[ht]
\vskip 15 truecm
\includegraphics{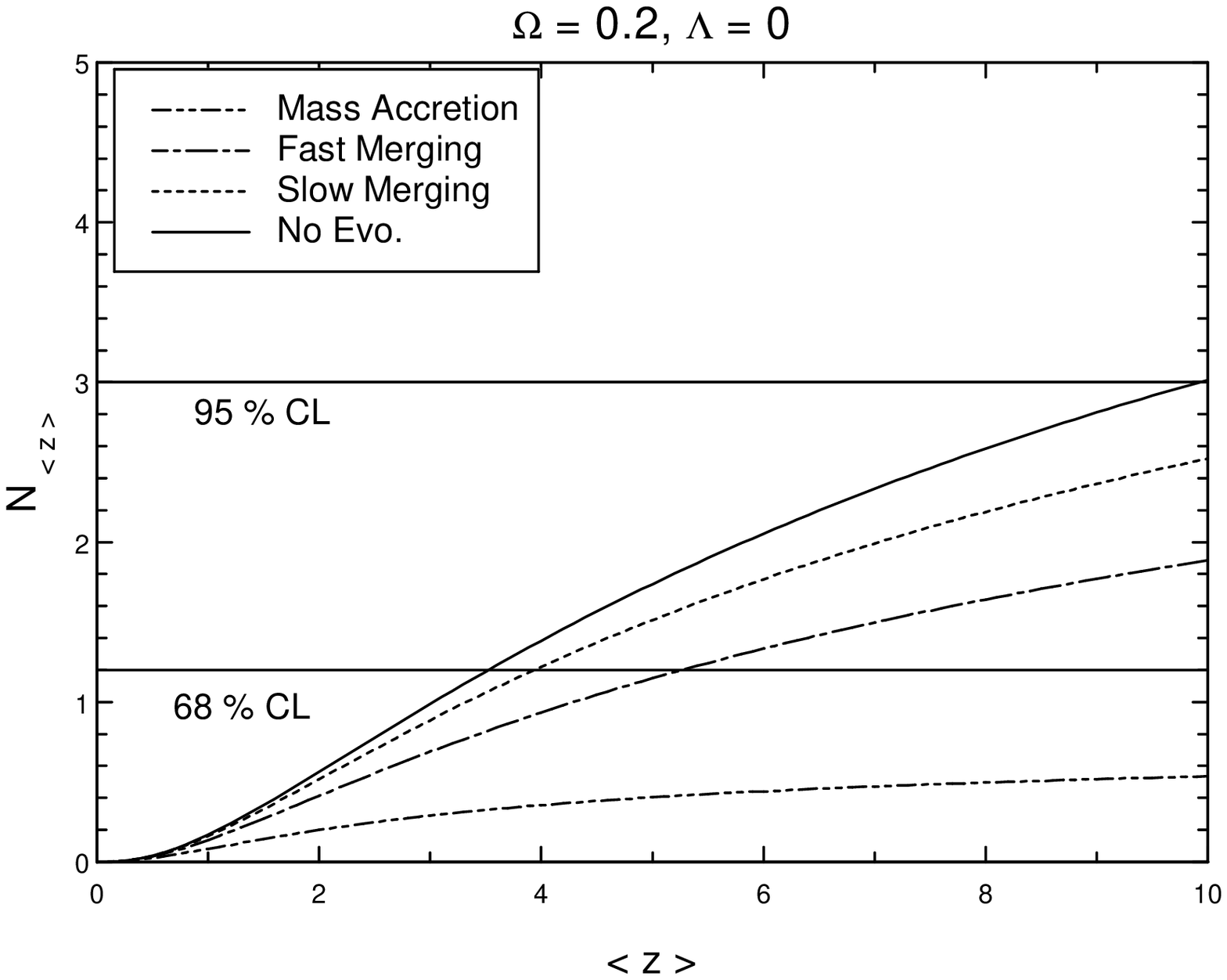}
\caption{}
\end{figure}

\begin{section}*{Figure Captions}
{\bf{Figure 1}}.  Expected number of observable image pairs in the BATSE 4B
($N_{< z >}$) as function of $ < z > $, the effective average redshift of
GRBs, in flat universe with $\Omega = 0.3 $ and $\Lambda = 0.7$ .
\vskip 0.2in
\noindent
{\bf{Figure 2}}.  Expected number of observable image pairs in the BATSE 4B
($N_{< z >}$) as function of $ < z > $, the effective average redshift of
GRBs, in flat universe with $\Omega = 0.5 $ and $\Lambda = 0.5$ .
\vskip 0.2in
\noindent
{\bf{Figure 3}}.  Expected number of observable image pairs in the BATSE 4B
($N_{< z >}$) as function of $ < z > $, the effective average redshift of
GRBs, in an open universe with $\Omega = 0.2 $ and $\Lambda = 0.0$ .

   \end {section}
   \vfill
   \eject

\begin{table}
\title{Table 1. Limits On The Average Redshift $\left < z \right>$ Of
GRBs In Different Models Of Galaxy Evolution At 95$\%$ Confidence
Level with 1802 bursts} 
\begin{center}
\begin{tabular}{l|llll}\hline\hline
$\Omega + \Lambda$ & $ No \ Evo.$ & $ Slow \ Merging$ & $ Fast \ Merging$
& $ Mass\  Accretion$  \\   \hline\hline
$0.3 + 0.7$ & $4.3$ &$5.0$ &$7.6$ &$--$ \\  \\  
$0.5 + 0.5$ & $7.8$ & $10.0$ &$--$ &$--$   \\ \\
$0.2 + 0.0$ & $9.8$ & $--$ & $--$ &$--$   \\ \\
\hline
\end{tabular}

\end{center}
\end{table}

\begin{table}
\title{Table 2. Limits On The Average Redshift $\left < z \right>$ Of
GRBs In  Different Models Of Galaxy Evolution At 68$\%$ Confidence
Level with  1802 bursts}
\begin{center}
\begin{tabular}{l|llll}\hline\hline
$\Omega + \Lambda$ & $ No \ Evo.$ & $ Slow\ Merging$ & $ Fast\
Merging$ & $ Mass\ Accretion$  \\   \hline\hline
$0.3 + 0.7$ & $2.2$ &$2.3$ &$2.9$ &$6.1$ \\  \\  
$0.5 + 0.5$ & $3.0$ & $3.3$ &$4.3$ &$--$   \\ \\
$0.2 + 0.0$ & $3.5$ & $3.9$ & $5.2$ &$--$   \\ \\
\hline
\end{tabular}

\end{center}
\end{table}

\end{document}